\documentclass[11pt,a4paper]{article}
\pdfoutput=1

\usepackage{jheppub}
\usepackage{amsmath}
\usepackage{amssymb}
\usepackage{graphics}
\usepackage[active]{srcltx}
\usepackage{pdfsync}
\usepackage{shuffle}
\usepackage{slashed}
\usepackage{hyperref}
\usepackage{subfigure}

\newcommand{\bea}{\begin{eqnarray}}
\newcommand{\eea}{\end{eqnarray}}
\newcommand{\nn}{\nonumber \\}

\def\W #1{\widetilde{#1}}

\def\eref#1{(\ref{#1})}

\def\a{{\alpha}}

\def\b{{\beta}}

\def\la{\lambda}

\allowdisplaybreaks

\title{Can Locality, Unitarity, and Hidden Zeros Completely Determine Tree-Level Amplitudes?}
\author[a]{Kang Zhou}

\affiliation[a]{Center for Gravitation and Cosmology, College of Physical Science and Technology, Yangzhou University,\\
No.180, Siwangting Road, Yangzhou, 225009, P.R. China}

\emailAdd{zhoukang@yzu.edu.cn}

\date{\today}
\abstract{In this note, we address the question of whether locality, unitarity, and newly discovered hidden zeros can completely determine tree-level amplitudes, from the perspective of soft limit. We reconstruct the single-soft theorems of tree YM amplitudes and the double-soft theorems of tree NLSM amplitudes from locality, unitarity, and hidden zeros. A series of studies have shown that the full YM and NLSM amplitudes can be constructed from these soft theorems; therefore, we conclude that locality, unitarity, and hidden zeros completely determine the tree-level YM and NLSM amplitudes.
}


\begin{document}

\maketitle \flushbottom

\section{Introduction}
\label{sec-intro}

In the modern study of scattering amplitudes, a milestone idea is to directly construct amplitudes based on physical criteria, without relying on traditional Lagrangians or Feynman rules. A paradigmatic example is the BCFW on-shell recursion relation at tree-level \cite{Britto:2004ap,Britto:2005fq}. This approach, rooted in locality and unitarity, recursively constructs tree-level amplitudes through the factorization behavior of residues at physical poles.

However, relying solely on factorization at poles is usually insufficient to construct the complete amplitude. For instance, when applying the BCFW method to Yang-Mills (YM) amplitudes, one relies on gauge invariance to ensure that the boundary terms can be avoided by choosing appropriate BCFW-deformations \cite{Arkani-Hamed:2008bsc}. This indicates that the foundation underlying the on-shell calculation of tree YM amplitudes is that locality, unitarity, together with gauge invariance, completely determine the amplitude.
There is also another scenario where, although boundary terms cannot be avoided, the on-shell method can be generalized by incorporating new criteria. For example, combining the BCFW method with the Adler zero leads to new recursion relations applicable to a class of effective field theories, exemplified by the nonlinear sigma model (NLSM) \cite{Cheung:2015ota}. Evidently, the foundation underlying this new approach is that locality, unitarity, together with the Adler zero, completely determine the amplitudes.

From the above discussion, it is evident that obtaining new on-shell methods that do not rely on Lagrangians or Feynman rules requires specifying a minimal set of criteria sufficient to determine the amplitudes. More precisely, we need a set $\{$locality, unitarity, $X\}$. For YM amplitudes, $X$ is gauge invariance; for NLSM amplitudes, $X$ is the Adler zero. However, a less satisfactory aspect emerging from the above two examples is that each physical model appears to require its own specific $X$.
Thus, a natural question arises: does there exist a universal $X$ that applies to a broad class of models?

The hidden zeros of tree-level amplitudes, discovered recently, hold promise for answering the above question. In \cite{Arkani-Hamed:2023swr}, the authors found that when the kinematic variables of external particles satisfy certain constraints, the tree-level amplitudes for ${\rm Tr}(\phi^3)$, NLSM, and YM theories vanish. This behavior is referred to as hidden zeros. In subsequent works, hidden zeros were extended to the tree-level amplitudes of a broader range of physical models, including Einstein gravity (GR), Dirac-Born-Infeld, special Galileon, as well as Yang-Mills and GR amplitudes with specific higher derivative corrections \cite{Rodina:2024yfc,Bartsch:2024amu,Li:2024qfp,Zhang:2024efe,Huang:2025blb,Zhou:2025tvq}. The zeros of loop-level Feynman integrands have also been investigated accordingly \cite{Backus:2025hpn}. Prior to this, the study of zeros of amplitudes had been far less extensive and in-depth than that of poles. The aforementioned developments have partially filled this gap.

Since the hidden zero is a novel analytic property distinct from the factorization at poles, one may naturally ask that whether hidden zeros could serve as the $X$ in the set $\{$locality, unitarity, $X\}$. If the answer is affirmative, then we obtain a universal $X$ that covers a wide range of models. The most direct approach to addressing this question is to establish an effective method based on locality, unitarity, and hidden zeros, capable of computing amplitudes. However, this is not an easy task. In previous work \cite{Li:2025suo}, we combined the BCFW approach with hidden zeros and developed an on-shell recursive method applicable to NLSM amplitudes. Unfortunately, this method is too specific, and at present, it is entirely unclear how to extend it to other models such as YM.

Given the above difficulty, in this note we approach the question of whether hidden zeros can play the role of $X$ from another, less direct perspective. Various research in the area of scattering amplitudes has shown that a large variety of tree amplitudes can be completely determined by their behavior in the soft limit \cite{Elvang:2018dco,Nguyen:2009jk,Boucher-Veronneau:2011rwd,Rodina:2018pcb,Ma:2022qja,Luo:2015tat,Zhou:2022orv,
Du:2024dwm,Zhou:2024qjh,Zhou:2024qwm,Wei:2023yfy,Hu:2023lso}. Therefore, we can investigate whether locality, unitarity, together with hidden zeros, can fully determine such soft behaviors. The answer to this new question will directly reveal whether hidden zeros can serve as a universal $X$.

Therefore, in this note, we construct the soft behaviors of YM and NLSM amplitudes based on locality, unitarity, and hidden zeros. More specifically, for YM, we construct the single-soft behavior obtained by taking the soft limit of a single gluon; for NLSM, we construct the double-soft behavior obtained by taking the simultaneous soft limit of two pions. We determine the parts of the soft behavior that contain specific poles through factorization at these poles, as dictated by locality and unitarity, and then probe the parts without such poles using hidden zeros. In the absence of any known theorem-such as one stating that locality, unitarity, together with hidden zeros, completely determine the amplitude-we cannot logically guarantee that the soft behavior constructed through the above approach is complete. Nevertheless, we can compare the results obtained from the above approach with known soft theorems. Since the soft limit behaviors of the YM and NLSM amplitudes obtained through the above method exactly match the known standard soft theorems, we conclude that locality, unitarity, and hidden zeros completely determine the YM and NLSM amplitudes.

Near each hidden zero, there exists a novel factorization behavior known as the $2$-split \cite{Cao:2024gln,Cao:2024qpp,Arkani-Hamed:2024fyd,Guevara:2024nxd,Zhou:2024ddy,Feng:2025ofq,Feng:2025dci,Cao:2025hio,Zhang:2025zjx,Zhang:2026dcm}. In previous work \cite{Zhou:2025xly}, we combined the traditional factorization at poles with the new factorization near zeros to construct the soft behaviors of amplitudes. It is therefore worthwhile to compare the method developed in this note with that previous approach. Since each $2$-split gives rise to two off-shell currents, constructing soft behavior via $2$-splits cannot rely solely on on-shell information; in contrast, the property of the hidden zero is entirely on-shell. Therefore, the construction of soft behavior presented in this note is significantly more convenient. Moreover, from the perspective of constructing amplitudes, the set $\{$locality, unitarity, hidden zeros$\}$ is clearly more advantageous than the set $\{$locality, unitarity, $2$-splits$\}$. The reason is again that employing $2$-splits confronts one with off-shell currents, which, due to being off-shell, also cause a host of complexities such as dependence on the choice of gauge.

The remainder of this note is organized as follows. In section \ref{sec-zero}, we review the hidden zeros of tree YM and NLSM amplitudes.
In section \ref{sec-YM}, we construct the single-soft theorems of tree YM amplitudes at leading and sub-leading orders, by utilizing locality, unitarity and hidden zeros. We also check the consistency from two aspects. Then, in section \ref{sec-NLSM} we construct the double-soft theorems of tree NLSM amplitudes at leading and sub-leading orders, by applying the same method, and also verify the consistency. Finally, we end with a brief conclusion and discussion in section \ref{sec-conclu}.

\section{Hidden zeros of ordered tree YM and NLSM amplitudes}
\label{sec-zero}

For readers' convenience, in this section we briefly review the hidden zeros of ordered YM and NLSM amplitudes at the tree-level.

\begin{figure}
  \centering
   \includegraphics[width=6cm]{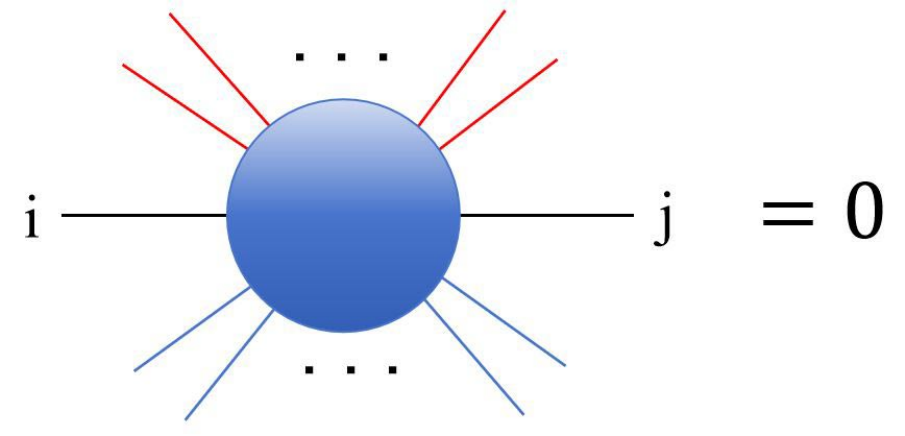} \\
  \caption{Diagrammatical illustration for the hidden zero. The red lines represent external particles in $\pmb A$, while blue lines represent external particles in $\pmb B$.}\label{zero}
\end{figure}

The hidden zeros for a given amplitude do not exhaust all zeros but instead represent a proper subset thereof. We first introduce such zeros for color-ordered YM amplitudes. Without loss of generality, we shall consider the $n$-point YM amplitude ${\cal A}_{\rm YM}(1,\cdots,n)$ with the canonical ordering. Each hidden zero of this amplitude can be achieved as follows. One can choose two external gluons, $i$ and $j$, that are non-adjacent in the ordering. The gluons between $i$ and $j$ are then grouped into two ordered sets $\pmb A=\{i+1,\cdots,j-1\}$ and $\pmb B=\{j+1,\cdots,i-1\}$. The hidden zero then states that
\bea
{\cal A}_{\rm YM}(1,\cdots,n)=0\,,~~~~{\rm if}~\{\epsilon_a\,,\,k_a\}\cdot\{\epsilon_b\,,\,k_b\}=0\,,~{\rm for}~\forall\,a\in\pmb A\,,\,b\in\pmb B\,.~~\label{zero-YM}
\eea
In the above, the notation $\{\epsilon_a\,,\,k_a\}\cdot\{\epsilon_b\,,\,k_b\}=0$ is understood as
\bea
\epsilon_a\cdot\epsilon_b=\epsilon_a\cdot k_b=k_a\cdot\epsilon_b=k_a\cdot k_b=0\,.
\eea
Such behavior is illustrated in Fig.\ref{zero}.
We emphasize that each hidden zero is associated with a specific choice of the pair $i$ and $j$. In the rest of this note, we denote this pair simply as $(i,j)$.

As a simple example, one can consider the $4$-point amplitude ${\cal A}_{\rm YM}(1,2,3,4)$, choose $(i,j)=(1,3)$, $\pmb A=\{2\}$, $\pmb B=\{4\}$,
and verify
\bea
{\cal A}_{\rm YM}(1,2,3,4)=0\,,~~~~{\rm if}~\{\epsilon_2\,,\,k_2\}\cdot\{\epsilon_4\,,\,k_4\}=0\,.
\eea

The hidden zeros for flavor-ordered NLSM amplitudes are similar. For a given $n$-point amplitude ${\cal A}_{\rm NLSM}(1,\cdots,n)$, where $n$ is an even number, one can also choose a pair of pions $(i,j)$, and denote pions between them as $\pmb A=\{i+1,\cdots,j-1\}$ and $\pmb B=\{j+1,\cdots,i-1\}$. The hidden zero associated with the pair $(i,j)$ is given as
\bea
{\cal A}_{\rm NLSM}(1,\cdots,n)=0\,,~~~~{\rm if}~k_a\cdot k_b=0\,,~{\rm for}~\forall\,a\in\pmb A\,,\,b\in\pmb B\,.~~\label{zero-NLSM}
\eea

In subsequent sections, we will use the hidden zeros of YM and NLSM amplitudes in \eref{zero-YM} and \eref{zero-NLSM}, together with locality and unitarity, to construct soft theorems of these amplitudes.

\section{Single-soft theorems of YM amplitudes from locality, unitarity and hidden zeros}
\label{sec-YM}

In this section, we construct the leading and sub-leading single-soft theorems of color-ordered YM amplitudes at tree-level, by using locality, unitarity and hidden zeros.

The single-soft limit for a massless external particle can be achieved as follows.
We parameterize the momentum of the soft particle as $k_s\to\tau\hat{k}_s$, and take the limit $\tau\to0$. The different orders are defined by the powers of the parameter $\tau$. The soft theorem at $i^{\rm th}$ order states that the amplitude at this order behaves as
\bea
{\cal A}^{(i)}_{n+1}={\cal S}^{(i)}\,{\cal A}_n\,,
\eea
where ${\cal A}^{(i)}_{n+1}$ is the $n+1$-point amplitude including the soft particle $s$, while ${\cal A}_n$ is the corresponding $n$-point amplitude with the soft particle $s$ removed. The factor ${\cal S}^{(i)}$ is called the soft factor at $i^{\rm th}$ order, and often behaves as an operator acting on ${\cal A}_n$. As well known, for YM amplitudes under consideration, the above form of soft theorem exists at the leading and sub-leading orders \cite{Casali:2014xpa}. That is,
\bea
{\cal A}^{(i)}_{\rm YM}(s,1,\cdots,n)={\cal S}^{(i)}_{\rm YM}(n,s,1)\,{\cal A}_{\rm YM}(1,\cdots,n)\,,
\eea
where $i=0,1$.

As studied in \cite{Rodina:2018pcb,Zhou:2022orv,Hu:2023lso}, the tree YM amplitudes can be completely determined by imposing their leading and sub-leading soft theorems. Thus, for our purpose, it is sufficient to construct the above two soft theorems from locality, unitarity and hidden zeros, without worrying about higher orders. In subsection \ref{subsec-YM-leading}, we review that the leading soft theorem can be determined by employing solely locality and unitarity.
Then, in subsection \ref{subsec-YM-subleading}, we construct the sub-leading soft theorem by utilizing locality, unitarity, as well as the hidden zero with a special choice of the pair $(i,j)$. Finally, in subsection \ref{subsec-YM-verify}, we verify the consistency of our sub-leading result from two aspects.

\subsection{Leading order}
\label{subsec-YM-leading}

As is well known, the leading soft theorem of tree YM amplitudes can be completely determined via locality and unitarity. In this subsection, we review such a construction for the leading soft theorem, and introduce notations which will be used frequently in the rest of this note.

\begin{figure}
  \centering
   \includegraphics[width=4cm]{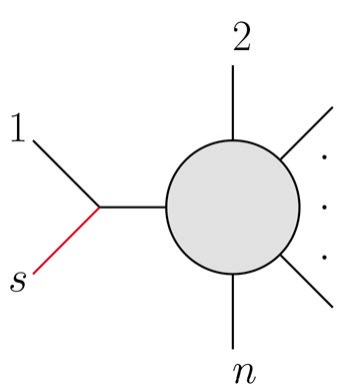}
   ~~~~~~~~~~~~~~~~~~~~
   \includegraphics[width=4cm]{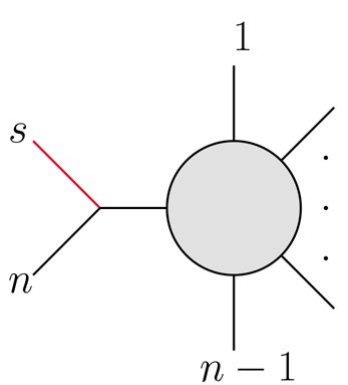} \\
  \caption{Diagrams with divergent propagators in the single-soft limit of YM amplitude, where the soft gluon represented in red is attached to the cubic vertex in each diagram.}\label{pole-part}
\end{figure}

The leading soft behavior obviously arises from $2$-point channels in Fig.\ref{pole-part} which cause divergent propagators $1/s_{s1}$ and $1/s_{ns}$, since $s_{s1}=2\tau k_s\cdot k_1$ and $s_{ns}=2\tau k_s\cdot k_n$. Thus, we can divide the $n+1$-point amplitude
${\cal A}_{\rm YM}(s,1,\cdots,n)$ into
\bea
{\cal A}_{\rm YM}(s,1,\cdots,n)={{\cal J}_{\rm YM}(s,1)\cdot{\cal J}_{\rm YM}(2,\cdots,n)\over s_{s1}}+{{\cal J}_{\rm YM}(n,s)\cdot{\cal J}_{\rm YM}(1,\cdots,n-1)\over s_{ns}}+{\cal R}\,,~~\label{divide-AYM}
\eea
where ${\cal J}^\mu_{\rm YM}$ are vector currents, and ${\cal R}$ denotes the remaining part which do not contain the propagator $1/s_{s1}$ or $1/s_{ns}$. The leading soft behavior then reads
\bea
&&{\cal A}^{(0)}_{\rm YM}(s,1,\cdots,n)={{\cal J}^{(0)}_{\rm YM}(s,1)\cdot{\cal J}^{(0)}_{\rm YM}(2,\cdots,n)\over s_{s1}}+{{\cal J}^{(0)}_{\rm YM}(n,s)\cdot{\cal J}^{(0)}_{\rm YM}(1,\cdots,n-1)\over s_{ns}}\nn
&&~~~~~~~~={\sum_{h}\,{\cal A}_{\rm YM}(s,1,I^h_L)\,{\cal A}_{\rm YM}(I^{-h}_R,2,\cdots,n)\over s_{s1}}
+{\sum_{h}\,{\cal A}_{\rm YM}(n,s,I'^h_L)\,{\cal A}_{\rm YM}(1,\cdots,n-1,I'^{-h}_R)\over s_{ns}}\,,~~\label{YM-leading-middle}
\eea
where the following relation is implicitly used,
\bea
\sum_{h}\,\big(\epsilon^h\,\epsilon^{-h}\big)^{\mu\nu}\,\sim\,g^{\mu\nu}\,.~~\label{polari-metric}
\eea
In \eref{YM-leading-middle}, ${\cal A}_{\rm YM}$ appear in numerators are on-shell amplitudes, due to the following reason. Momentum conservation forces
\bea
-k_{I_L}=k_{I_R}=k_s+k_1=\tau\hat{k}_s+k_1\,,~~~~-k_{I'_L}=k_{I'_R}=k_s+k_n=\tau\hat{k}_s+k_n\,.~~\label{relation-momen}
\eea
At the leading order, we have
\bea
k_s=0\,,~~~~-k^{(0)}_{I_L}=k^{(0)}_{I_R}=k_1\,,~~~~-k^{(0)}_{I'_L}=k^{(0)}_{I'_R}=k_n\,,~~\label{YM-k-leading}
\eea
thus all external momenta of each ${\cal A}_{\rm YM}$ are on-shell. Then the structure in \eref{YM-leading-middle} follows from the well-known factorization behavior on poles, which itself originates from locality and unitarity.

To construct the leading soft theorem using only locality and unitarity, the $3$-point amplitudes in \eref{YM-leading-middle} should be determined via bootstrap, not from Feynman rules. The process is simple: for a given ${\cal A}_{\rm YM}(a,b,c)$, the linearity in each polarization, the cyclical invariance and the mass dimension requires
\bea
{\cal A}_{\rm YM}(a,b,c)=(\epsilon_a\cdot\epsilon_b)\,\epsilon_c\cdot(p\,k_a+q\,k_b)+{\rm cyclic}\,,
\eea
where $p$ and $q$ are two constants. Using momentum conservation, as well as the on-shell condition $\epsilon_c\cdot k_c=0$, one can remove one of $p\,k_a$ and $q\,k_b$.
For latter convenience, we fix it as
\bea
{\cal A}_{\rm YM}(a,b,c)=(\epsilon_a\cdot k_c)\,(\epsilon_b\cdot\epsilon_c)+(\epsilon_b\cdot k_a)\,(\epsilon_c\cdot\epsilon_a)+(\epsilon_c\cdot k_b)\,(\epsilon_a\cdot\epsilon_b)\,,~~\label{YM-3pt}
\eea
up to an overall constant.

From \eref{YM-3pt}, we see that the $3$-point amplitudes in \eref{YM-leading-middle}, satisfying the configuration of momenta in \eref{YM-k-leading}, are given by
\bea
{\cal A}_{\rm YM}(s,1,I^h_L)&=&-(\epsilon_s\cdot k_1)\,(\epsilon_1\cdot\epsilon_{I_L}^h)\nn
{\cal A}_{\rm YM}(n,s,{I'_L}^h)&=&(\epsilon_s\cdot k_n)\,(\epsilon_n\cdot\epsilon_{I'_L}^h)\,.~~\label{3pt-YM-leading}
\eea
Plugging them into \eref{YM-leading-middle}, and using \eref{polari-metric}, as well as
\bea
{\cal A}_{\rm YM}(I^{-h}_R,2,\cdots,n)&=&{\cal A}_{\rm YM}(1,\cdots,n)\big|_{\epsilon_1\to\epsilon^{-h}_{I_R}}\,,\nn
{\cal A}_{\rm YM}(1,\cdots,n-1,I'^{-h}_R)&=&{\cal A}_{\rm YM}(1,\cdots,n)\big|_{\epsilon_n\to\epsilon^{-h}_{I'_R}}\,,~~\label{npt-YM-leading}
\eea
we find the standard leading soft theorem \cite{Casali:2014xpa}
\bea
{\cal A}^{(0)}_{\rm YM}(s,1,\cdots,n)={\cal S}^{(0)}_{\rm YM}(n,s,1)\,{\cal A}_{\rm YM}(1,\cdots,n)\,,
\eea
where the leading soft factor is given as
\bea
{\cal S}^{(0)}_{\rm YM}(n,s,1)={\epsilon_s\cdot k_n\over s_{ns}}-{\epsilon_s\cdot k_1\over s_{s1}}={1\over\tau}\,\Big({\epsilon_s\cdot k_n\over 2\,\hat{k}_s\cdot k_n}-{\epsilon_s\cdot k_1\over 2\,\hat{k}_s\cdot k_1}\Big)\,.
\eea

Clearly, the above derivation relies solely on locality and unitarity, without exploiting other properties such as gauge invariance or hidden zeros.
The reason behind this is that every term in the leading soft factor contains a pole in the soft limit. When we consider the sub-leading soft behavior in the next subsection, we will encounter contributions that do not contain these poles. Consequently, the sub-leading soft factor cannot be completely reconstructed from locality and unitarity.

\subsection{Sub-leading order}
\label{subsec-YM-subleading}

Now we consider the sub-leading order in $\tau$. Using the separation \eref{divide-AYM}, we see that the sub-leading soft behavior includes the following blocks,
\bea
{\cal A}^{(1)}_{\rm YM}(s,1,\cdots,n)&=&{{\cal J}^{(1)}_{\rm YM}(s,1)\cdot{\cal J}^{(0)}_{\rm YM}(2,\cdots,n)\over s_{s1}}+{{\cal J}^{(1)}_{\rm YM}(n,s)\cdot{\cal J}^{(0)}_{\rm YM}(1,\cdots,n-1)\over s_{ns}}\nn
&&+{{\cal J}^{(0)}_{\rm YM}(s,1)\cdot{\cal J}^{(1)}_{\rm YM}(2,\cdots,n)\over s_{s1}}+{{\cal J}^{(0)}_{\rm YM}(n,s)\cdot{\cal J}^{(1)}_{\rm YM}(1,\cdots,n-1)\over s_{ns}}\nn
&&+{\cal R}^{(0)}\,.~~\label{YM-subleading-blocks}
\eea
Evidently, locality and unitarity can not determine the block ${\cal R}^{(0)}$. We will show that this part can be fixed by imposing the hidden zero
with $(i,j)=(1,n)$.

Before dealing with ${\cal R}^{(0)}$, let us try to figure out the other blocks. At the leading order, we found
\bea
&&{{\cal J}^{(0)}_{\rm YM}(s,1)\cdot{\cal J}^{(0)}_{\rm YM}(2,\cdots,n)\over s_{s1}}+{{\cal J}^{(0)}_{\rm YM}(n,s)\cdot{\cal J}^{(0)}_{\rm YM}(1,\cdots,n-1)\over s_{ns}}\nn
&=&{\sum_{h}\,{\cal A}_{\rm YM}(s,1,I^h_L)\,{\cal A}_{\rm YM}(I^{-h}_R,2,\cdots,n)\over s_{s1}}
+{\sum_{h}\,{\cal A}_{\rm YM}(n,s,I'^h_L)\,{\cal A}_{\rm YM}(1,\cdots,n-1,I'^{-h}_R)\over s_{ns}}\,,~~\label{leading-forsub}
\eea
where the external momenta of these ${\cal A}_{\rm YM}$ are constrained by \eref{YM-k-leading}. At the sub-leading order, we should replace the momenta configuration with the more general one \eref{relation-momen} to account for $k_s$. Such replacement allows us to find pieces contribute to the first and second lines on the r.h.s of \eref{YM-subleading-blocks}.

We first do such a replacement to $3$-point amplitudes in \eref{leading-forsub}. Combining the momenta configuration in \eref{relation-momen} and the general form of the $3$-point amplitude in \eref{YM-3pt}, we get:
\bea
&&\sum_{h}\,A^{(1)}_{\rm YM}(s,1,I^h_L)\,{\cal A}_{\rm YM}(I^{-h}_R,2,\cdots,n)\in{\cal J}^{(1)}_{\rm YM}(s,1)\cdot{\cal J}^{(0)}_{\rm YM}(2,\cdots,n)\,,\nn
&&\sum_{h}\,A^{(1)}_{\rm YM}(n,s,I'^h_L)\,{\cal A}_{\rm YM}(1,\cdots,n-1,I'^{-h}_R)\in{\cal J}^{(1)}_{\rm YM}(n,s)\cdot{\cal J}^{(0)}_{\rm YM}(1,\cdots,n-1)\,,~~\label{YM-sub-piece1}
\eea
where
\bea
A^{(1)}_{\rm YM}(s,1,I^h_L)&=&\tau\,\epsilon_1\cdot \hat{f}_{s}\cdot\epsilon^h_{I_L}\,,\nn
A^{(1)}_{\rm YM}(n,s,I'^h_L)&=&-\tau\,\epsilon_n\cdot\hat{f}_s\cdot\epsilon^h_{I'_L}\,.~~\label{3pt-YM-subleading}
\eea
In the above, the strength tensor is defined as $\hat{f}^{\mu\nu}_s\equiv\hat{k}_s^\mu\epsilon_s^\nu-\epsilon_s^\mu\hat{k}_s^\nu$.
We use the notation $A^{(1)}_{\rm YM}$ instead of ${\cal A}^{(1)}_{\rm YM}$ to highlight that these ingredients serve as sub-leading corrections to the corresponding on-shell amplitudes at leading order.
Given that we confine ourselves to locality and unitarity alone, without invoking additional tools such as Feynman rules or gauge invariance, there is no logical basis to conclude that the above derivation exactly yields the entire first line on the r.h.s. of \eref{YM-subleading-blocks}. This is reason why we use $\in$ in \eref{YM-sub-piece1}.

Meanwhile, applying such a replacement to ${\cal A}_{\rm YM}(I^{-h}_R,2,\cdots,n)$ and ${\cal A}_{\rm YM}(1,\cdots,n-1,I'^{-h}_R)$ leads to
\bea
&&\sum_{h}\,{\cal A}_{\rm YM}(s,1,I^h_L)\,A^{(1)}_{\rm YM}(I^{-h}_R,2,\cdots,n)\in{\cal J}^{(0)}_{\rm YM}(s,1)\cdot{\cal J}^{(1)}_{\rm YM}(2,\cdots,n)\,,\nn
&&\sum_{h}\,{\cal A}_{\rm YM}(n,s,I'^h_L)\,A^{(1)}_{\rm YM}(1,\cdots,n-1,I'^{-h}_R)\in{\cal J}^{(0)}_{\rm YM}(n,s)\cdot{\cal J}^{(1)}_{\rm YM}(1,\cdots,n-1)\,,~~\label{YM-sub-piece2}
\eea
where
\bea
A^{(1)}_{\rm YM}(I^{-h}_R,2,\cdots,n)&=&\tau\,\hat{k}_s\cdot\partial_{k_1}{\cal A}_{\rm YM}(1,\cdots,n)\big|_{\epsilon_1\to\epsilon^{-h}_{I_R}}\,,\nn
A^{(1)}_{\rm YM}(1,\cdots,n-1,I'^{-h}_R)&=&\tau\,\hat{k}_s\cdot\partial_{k_n}{\cal A}_{\rm YM}(1,\cdots,n)\big|_{\epsilon_n\to\epsilon^{-h}_{I'_R}}\,.~~\label{npt-YM-subleading}
\eea
The effect of the operator $\hat{k}_s\cdot\partial_{k_1}$ is replacing $k_1$ in ${\cal A}_{\rm YM}(1,\cdots,n)$ with $k_1+\tau\hat{k}_s$ and expanding it to the sub-leading order in $\tau$. The analogous interpretation holds for $\hat{k}_s\cdot\partial_{k_n}$.
Notice that the operators $\hat{k}_s\cdot\partial_{k_1}$ and $\hat{k}_s\cdot\partial_{k_n}$ are individually inconsistent with momentum conservation. For instance, the first operator does not make sense if one use momentum conservation to eliminate $k_1$ in the expression of
${\cal A}_{\rm YM}(1,\cdots,n)$. However, in the next subsection we will show that such inconsistency does not appear in the final combinatorial operator \eref{soft-factor-YMsub} which gives rise to the sub-leading soft behavior.

By plugging \eref{YM-sub-piece1} and \eref{YM-sub-piece2} into \eref{YM-subleading-blocks}, and utilizing the relation \eref{polari-metric}, we get
\bea
{\cal A}^{(1)}_{\rm YM}(s,1,\cdots,n)
&=&\Big[{\epsilon_1\cdot\hat{f}_s\cdot\partial_{\epsilon_1}-(k_1\cdot\epsilon_s)(\hat{k}_s\cdot\partial_{k_1})\over 2\,\hat{k}_s\cdot k_1}-{\epsilon_n\cdot\hat{f}_s\cdot\partial_{\epsilon_n}-(k_n\cdot\epsilon_s)(\hat{k}_s\cdot\partial_{k_n})\over 2\,\hat{k}_s\cdot k_n}\Big]\,{\cal A}_{\rm YM}(1,\cdots,n)\nn
&&+{\cal M}+{\cal R}^{(0)}\,,~~\label{YM-subleading-middle}
\eea
where ${\cal M}$ denotes the potential missed terms from the first and second lines on the r.h.s of \eref{YM-subleading-blocks}. For simplicity, we now define ${\cal R}'={\cal M}+{\cal R}^{(0)}$.

Our next task is to find ${\cal R}'$. Before delving into the details, we first observe that ${\cal R}'$ does not have a pole at $s_{ns}=0$ or $s_{s1}=0$.
To see this, consider the case that $\tau$ is small but not infinitesimally small, and the pole at $s_{ns}=0$ is approached via a new path instead of by taking the soft limit $\tau\to0$. Clearly, in this case all $A^{(1)}_{\rm YM}$ in \eref{3pt-YM-subleading} and \eref{npt-YM-subleading} contribute to $3$-point and $n$-point on-shell amplitudes when evaluating the residue around this pole. However, such on-shell amplitudes which determine the residue, receive no contribution from ${\cal M}$; that is, ${\cal M}$ does not have a pole at $s_{ns}=0$. The analogous argument holds for $s_{s1}=0$. Consequently, although ${\cal M}$ arises from the first and second lines on the r.h.s of \eref{YM-subleading-blocks}, its denominators $s_{ns}$ and $s_{s1}$ are canceled by the numerators (if the non-zero ${\cal M}$ does exist). By definition, ${\cal R}^{(0)}$ also does not contain these poles.
Thus we conclude that ${\cal R}'$ does not have these poles.

To determine ${\cal R}'$, we impose the hidden zero with $(i,j)=(1,n)$. This zero states that the full amplitude vanishes if
\bea
\{\epsilon_s\,,\,k_s\}\cdot\{\epsilon_a\,,\,k_a\}=0\,,~~~~{\rm for}~\forall a\in\{2,\cdots,n-1\}\,.~~\label{kinematic-forYMsubleading}
\eea
Such behavior implies that under the zero kinematics ${\cal R}'$ should cancel the first line on the r.h.s of \eref{YM-subleading-middle}.

Under the kinematic condition \eref{kinematic-forYMsubleading}, the first useful observation is,
\bea
k_s\cdot(k_1+k_n)=-k_s\cdot(k_2+\cdots+k_{n-1})=0\,,
\eea
which implies $\hat{k}_s\cdot k_1=-\hat{k}_s\cdot k_n$. Therefore, we can turn the first line on the r.h.s of \eref{YM-subleading-middle} to
\bea
&&\Big[{\epsilon_1\cdot\hat{f}_s\cdot\partial_{\epsilon_1}-(k_1\cdot\epsilon_s)(\hat{k}_s\cdot\partial_{k_1})\over 2\,\hat{k}_s\cdot k_1}-{\epsilon_n\cdot\hat{f}_s\cdot\partial_{\epsilon_n}-(k_n\cdot\epsilon_s)(\hat{k}_s\cdot\partial_{k_n})\over 2\,\hat{k}_s\cdot k_n}\Big]\,{\cal A}_{\rm YM}(1,\cdots,n)\nn
&&\xrightarrow[]{\eref{kinematic-forYMsubleading}}\,{1\over 2\,\hat{k}_s\cdot k_1}\,\Big[\epsilon_1\cdot\hat{f}_s\cdot\partial_{\epsilon_1}+\epsilon_n\cdot\hat{f}_s\cdot\partial_{\epsilon_n}
-(k_1\cdot\epsilon_s)(\hat{k}_s\cdot\partial_{k_1})-(k_n\cdot\epsilon_s)(\hat{k}_s\cdot\partial_{k_n})\Big]\,{\cal A}_{\rm YM}(1,\cdots,n)\,.~~\label{impose0-step1-YM}
\eea
To proceed, we separate ${\cal A}_{\rm YM}(1,\cdots,n)$ as
\bea
{\cal A}_{\rm YM}(1,\cdots,n)=(\epsilon_1\cdot \epsilon_n)\,\a+(\epsilon_1\cdot k_n)\,\b+(\epsilon_n\cdot k_1)\,\gamma+\sum_{m\geq1}\,\lambda_m\,(k_1\cdot k_n)^m+\cdots\,.
\eea
The kinematic condition \eref{kinematic-forYMsubleading}, together with the linearity in each polarization and on-shell conditions $\{\epsilon_1,k_1\}\cdot k_1=0$, $\{\epsilon_n,k_n\}\cdot k_n=0$, leads to the following simple results
\bea
\epsilon_1\cdot\hat{f}_s\cdot\partial_{\epsilon_1}\,{\cal A}_{\rm YM}(1,\cdots,n)&=&(\epsilon_1\cdot\hat{f}_s\cdot\epsilon_n)\,\a+(\epsilon_1\cdot\hat{f}_s\cdot k_n)\,\b\,,\nn
\epsilon_n\cdot\hat{f}_s\cdot\partial_{\epsilon_n}\,{\cal A}_{\rm YM}(1,\cdots,n)&=&(\epsilon_n\cdot\hat{f}_s\cdot\epsilon_1)\,\a+(\epsilon_n\cdot\hat{f}_s\cdot k_1)\,\gamma\,,\nn
\hat{k}_s\cdot\partial_{k_1}\,{\cal A}_{\rm YM}(1,\cdots,n)&=&(\hat{k}_s\cdot\epsilon_n)\,\gamma+\sum_{m\geq1}\,m\lambda_m\,(\hat{k}_s\cdot k_n)\,(k_1\cdot k_n)^{m-1}\,,\nn
\hat{k}_s\cdot\partial_{k_n}\,{\cal A}_{\rm YM}(1,\cdots,n)&=&(\hat{k}_s\cdot\epsilon_1)\,\b+\sum_{m\geq1}\,m\lambda_m\,(\hat{k}_s\cdot k_1)\,(k_1\cdot k_n)^{m-1}\,.
\eea
Substituting these into \eref{impose0-step1-YM}, we arrive at
\bea
{\cal R}'\,\xrightarrow[]{\eref{kinematic-forYMsubleading}}&&\,{1\over 2\,\hat{k}_s\cdot k_1}\Big[(\epsilon_1\cdot\epsilon_s)(\hat{k}_s\cdot k_n)\,\b
+(\epsilon_n\cdot\epsilon_s)(\hat{k}_s\cdot k_1)\,\gamma\nn
&&+\Big((k_1\cdot\epsilon_s)(\hat{k}_s\cdot k_n)+(k_n\cdot\epsilon_s)(\hat{k}_s\cdot k_1)\Big)\,\Big(\sum_{m\geq1}\,m\lambda_m\,(k_1\cdot k_n)^{m-1}\Big)\Big]\,.~~\label{R0-form1}
\eea

Under the kinematic constraint \eref{kinematic-forYMsubleading}, it is straightforward to simplify \eref{R0-form1} as
\bea
{\cal R}'\,\xrightarrow[]{\eref{kinematic-forYMsubleading}}&&\,
{1\over2}\,\Big[(\epsilon_n\cdot\epsilon_s)\,\gamma-(\epsilon_1\cdot\epsilon_s)\,\b+\big(k_n\cdot\epsilon_s-k_1\cdot\epsilon_s\big)\,\Big(
\sum_{m\geq1}\,m\lambda_m\,(k_1\cdot k_n)^{m-1}\Big)\Big]\nn
=&&{1\over2}\,\Big(\epsilon_s\cdot\partial_{k_1}-\epsilon_s\cdot\partial_{k_n}\Big)\,{\cal A}_{\rm YM}(1,\cdots,n)\,,~~\label{R0-operator-YM}
\eea
where the observation $\hat{k}_s\cdot(k_1+k_n)=0$ is used again. Notice that this result does not contain $1/s_{s1}$ or $1/s_{ns}$, satisfying our expectation for ${\cal R}'$.
Plugging the formula in the second line of \eref{R0-operator-YM} into \eref{YM-subleading-middle}, we finally obtain
\bea
{\cal A}^{(1)}_{\rm YM}(s,1,\cdots,n)
&=&{\cal S}^{(1)}_{\rm YM}(n,s,1)\,{\cal A}_{\rm YM}(1,\cdots,n)\,+\,\W{\cal R}'\,,~~\label{YM-subleading-final}
\eea
where the soft factor is given as
\bea
{\cal S}^{(1)}_{\rm YM}(n,s,1)={\epsilon_1\cdot\hat{f}_s\cdot\partial_{\epsilon_1}+k_1\cdot\hat{f}_s\cdot\partial_{k_1}\over 2\,\hat{k}_s\cdot k_1}-{\epsilon_n\cdot\hat{f}_s\cdot\partial_{\epsilon_n}+k_n\cdot\hat{f}_s\cdot\partial_{k_n}\over 2\,\hat{k}_s\cdot k_n}\,.~~\label{soft-factor-YMsub}
\eea
The block $\W{\cal R}'$ collects potential terms in ${\cal R}'$ which vanish under the kinematic constraint \eref{kinematic-forYMsubleading}. In the above, we have used
\bea
\epsilon_s\cdot\partial_{k_1}={k_1\cdot \hat{k}_s\over k_1\cdot \hat{k}_s}\epsilon_s\cdot\partial_{k_1}\,,~~~~
\epsilon_s\cdot\partial_{k_n}={k_n\cdot \hat{k}_s\over k_n\cdot \hat{k}_s}\epsilon_s\cdot\partial_{k_n}\,,
\eea
to make the expression more compact.

As shown in \cite{Zhou:2022orv,Du:2024dwm}, the soft factor ${\cal S}^{(1)}_{\rm YM}(n,s,1)$ in \eref{soft-factor-YMsub} is equivalent to the standard one
\bea
{\cal S}^{(1)}_{\rm YM}(n,s,1)={\epsilon_s\cdot J_n\cdot \hat{k}_s\over 2\,\hat{k}_s\cdot k_n}-{\epsilon_s\cdot J_1\cdot \hat{k}_s\over 2\,\hat{k}_s\cdot k_1}\,,
\eea
where each $J_i$ is the angular momentum operator for the external particle $i$ \cite{Casali:2014xpa}. That is, $\W{\cal R}'=0$, thus we conclude that the sub-leading soft theorem of YM amplitudes is completely determined by imposing locality, unitarity and hidden zeros. It is worth emphasizing that the derivation in this subsection only gives rise to terms in ${\cal R}'$ which do not vanish under the constraint of zero kinematics. In other words, as discussed in section \ref{sec-intro}, one cannot logically conclude that $\W{\cal R}'=0$. However, our aim is to answer whether locality, unitarity and hidden zeros can fully fix the sub-leading soft theorem. For this purpose, it is sufficient to compare \eref{YM-subleading-final} with the known result and find wether or not $\W{\cal R}'=0$. As we have seen, such a comparison yields the desired conclusion.

On the other hand, although we cannot prove $\W{\cal R}'=0$ without comparing with known results, we can verify the consistency of
sub-leading soft factor in \eref{soft-factor-YMsub} from different perspectives. We will do such verifications in the next subsection.

\subsection{Consistency verification}
\label{subsec-YM-verify}

In this subsection, we will examine the self-consistency of the soft behavior in \eref{YM-subleading-final} and \eref{soft-factor-YMsub} (with $\W{\cal R}'=0$) from two aspects:
\begin{itemize}
\item[(1)] It satisfies hidden zeros with any choice of $(i,j)$.
\item[(2)] It is consistent with momentum conservation.
\end{itemize}

Let us begin with the first one.
That is, we verify wether or not the sub-leading soft behavior in \eref{YM-subleading-final}, obtained by imposing the hidden zero with $(i,j)=(1,n)$,
also exhibits all other hidden zeros for different choices of $(i,j)$. The verification is straightforward, as it is easy to see that the satisfaction of the zero kinematics for the $n+1$-point amplitude ${\cal A}_{\rm YM}(s,1,\cdots,n)$ implies that the zero kinematics for the
$n$-point amplitude ${\cal A}_{\rm YM}(1,\cdots,n)$ automatically holds for all cases except when $(i,j)=(1,n)$. To be more explicit, let us consider two examples. In the first example, we choose $(i,j)=(1,n-1)$ for both $n+1$-point amplitude and $n$-point amplitude. For the $n+1$-point amplitude, we have $\pmb A=\{2,\cdots,n-2\}$, $\pmb B=\{n,s\}$.
For the $n$-point amplitude, we have $\pmb A'=\{2,\cdots,n-2\}$, $\pmb B'=\{n\}$. We see that $\pmb A'=\pmb A$, $\pmb B'\subset\pmb B$, thus the zero kinematics of the $n+1$-point amplitude automatically ensures the zero kinematics of the $n$-point amplitude. In the second example, we choose $(i,j)=(s,n-2)$ for the $n+1$-point amplitude. This choice implies $\pmb A=\{1,\cdots,n-3\}$, $\pmb B=\{n-1,n\}$. Such $\pmb A$ and $\pmb B$ simultaneously ensure zero kinematics of $n$-point amplitude with two different choices of $(i,j)$, which are $(i,j)=(1,n-2)$ and $(i,j)=(n,n-2)$.
The first choice leads to $\pmb A'=\{2,\cdots,n-3\}$, $\pmb B'=\{n,n-1\}$, satisfying $\pmb A'\subset\pmb A$, $\pmb B'=\pmb B$. The second choice yields $\pmb A'=\{1,\cdots,n-3\}$, $\pmb B'=\{n-1\}$, satisfying $\pmb A'=\pmb A$, $\pmb B'\subset\pmb B$. Thus the zero kinematics of the $n$-point amplitude is automatically satisfied.

Therefore, imposing zero kinematics of $n+1$-point amplitude with $(i,j)\neq(1,n)$ yields ${\cal A}_{\rm YM}(1,\cdots,n)=0$. Meanwhile, since ordered sets $\pmb A$ and $\pmb B$ are always separated by $i$ and $j$, propagators $1/\hat{k}_s\cdot k_1$ and $1/\hat{k}_s\cdot k_n$ in \eref{YM-subleading-final} will never be divergent under any zero kinematics. Consequently, at the sub-leading order of $\tau$, hidden zeros of ${\cal A}^{(1)}_{\rm YM}(s,1,\cdots,n)$ with $(i,j)\neq(1,n)$ are guaranteed by hidden zeros of ${\cal A}_{\rm YM}(1,\cdots,n)$.

The second aspect for the YM case has been discussed in detail in our previous work \cite{Zhou:2025xly}; thus, here we only give a rapid review. We first clarify the meaning of ``consistent with momentum conservation". The soft factor in \eref{soft-factor-YMsub} can be understood as an operator acting on ${\cal A}_{\rm YM}(1,\cdots,n)$.
The expression of ${\cal A}_{\rm YM}(1,\cdots,n)$ is not unique; one can always freely modify it using momentum conservation, for example, by replacing $k_n$ with $-\sum_{i=1}^{n-1}k_i$. In general, changing the expression may alter the effect of the operator. When saying ``consistent with momentum conservation", we mean that if one changes the expression of the amplitude using momentum conservation, the effect of the operator remains unchanged.

For an $n$-point amplitude with the fixed number of external particles, a convenient way for verifying the above consistency is to define the momentum conservation operator \cite{Cheung:2017ems}
\bea
P_n=\sum_{i=1}^n\,K_i\cdot V\,,~~\label{defin-P}
\eea
where $V^\mu$ is a reference Lorentz vector, and $K_i^\mu$ are operators. When acting on a function ${\cal F}_n$ of $n$ external momenta, each $K_i$ takes the value of $k_i$ carried by ${\cal F}_n$. Suppose an operator ${\cal T}$ transmutes an $n$-point amplitude ${\cal A}_n$ to a function $G_n$, i.e., ${\cal T}\,{\cal A}_n=G_n$. Then $P_n{\cal A}_n=0$ and $P_n G_n=0$ imply the commutativity $[{\cal T}\,,\,P_n]\,{\cal A}_n=0$. Notice that we only require these two operators to commute effectively when acting on physical amplitudes ${\cal A}_n$. If this commutativity holds, then the action of ${\cal T}$ is independent of re-parameterizing ${\cal A}_n$ via momentum conservation.

As a simple example, consider the operator ${\cal T}=\partial_{\epsilon_a\cdot k_1}-\partial_{\epsilon_a\cdot k_n}$. This operator commutes with $P_n$ even when choosing $V=\epsilon_a$. It is straightforward to verify that the effect of this operator will not be altered if one re-parameterizes the amplitude using momentum conservation. For instance, if one replaces $k_n$ by $-\sum_{i=1}^{n-1}k_i$, the effect of $-\partial_{\epsilon_a\cdot k_n}$ will be compensated by $\partial_{\epsilon_a\cdot k_1}$.

For the soft theorem under consideration, the situation is slightly different, since the soft factor transmutes an $n$-point amplitude to an $n+1$-point one. Thus we need not only $P_n$ but also $P_{n+1}=P_n+K_s\cdot V$, where the operator $K_s$ takes the value of $k_s$ when acting on ${\cal A}_{\rm YM}(s,1,\cdots,n)$. Using
\bea
P_n\,{\cal A}_{\rm YM}(1,\cdots,n)=0\,,~~~~~~P_{n+1}\,{\cal A}_{\rm YM}(s,1,\cdots,n)=0\,,
\eea
one obtains the following consistency condition:
\bea
[{\cal S}^{(1)}_{\rm YM}(n,s,1)\,,\,P_n]\,{\cal A}_{\rm YM}(1,\cdots,n)=(k_s\cdot V)\,{\cal S}^{(0)}_{\rm YM}(n,s,1)\,{\cal A}_{\rm YM}(1,\cdots,n)\,.
\eea
The verification for this condition is straight forward, since acting ${\cal S}^{(1)}_{\rm YM}(n,s,1)$ on $P_n$ gives rise to
\bea
{\cal S}^{(1)}_{\rm YM}(n,s,1)\,P_n&=&{k_1\cdot\hat{f}_s\cdot V\over 2\,\hat{k}_s\cdot k_1}-{k_n\cdot\hat{f}_s\cdot V\over 2\,\hat{k}_s\cdot k_n}\nn
&=&\Big({\epsilon_s\cdot k_n\over 2\,\hat{k}_s\cdot k_n}-{\epsilon_s\cdot k_1\over 2\,\hat{k}_s\cdot k_1}\Big)\,(\hat{k}_s\cdot V)\,.
\eea
This means that the effect of acting ${\cal S}^{(1)}_{\rm YM}(n,s,1)$ on ${\cal A}_{\rm YM}(1,\cdots,n)$ is invariant under re-parameterizing ${\cal A}_{\rm YM}(1,\cdots,n)$ via momentum conservation.

\section{Double-soft theorems of NLSM amplitudes from locality, unitarity and hidden zeros}
\label{sec-NLSM}

In this section, we construct the leading and sub-leading double-soft theorems of ordered NLSM amplitudes at tree-level, by applying the same method based on locality, unitarity and hidden zeros, and verify the consistency from the same two aspects.

We label the two soft particles as $s_1$ and $s_2$, and parameterize their momenta as $k_{s_1}\to\tau\hat{k}_{s_1}$, $k_{s_2}\to\tau\hat{k}_{s_2}$.
Then the double-soft limit can be approached by taking $\tau\to 0$, which the same as the single-soft limit. We focus on the situation where $s_1$ and $s_2$ are adjacent to each other in the ordering. Thus, the goal of this section is to show
\bea
{\cal A}^{(i)}_{\rm NLSM}(s_1,s_2,1,\cdots,n)={\cal S}^{(i)}_{\rm NLSM}\,{\cal A}_{\rm NLSM}(1,\cdots,n)\,,
\eea
with $i=0,1$, and to find the explicit soft factors ${\cal S}^{(i)}_{\rm NLSM}$. Because we wish to develop a method of general applicability, the process in this section bears a strong resemblance to that in the previous section. Since the leading and sub-leading soft theorems are sufficient to determine the full NLSM amplitudes at the tree-level \cite{Rodina:2018pcb,Zhou:2024qjh}, we conclude that locality, unitarity and hidden zeros completely determine the tree NLSM amplitudes.

The leading and sub-leading double-soft theorems are derived in subsection \ref{subsec-NLSM-leading} and subsection \ref{subsec-NLSM-subleading}, respectively. The consistency of the result is verified in subsection \ref{subsec-NLSM-verify}.

\subsection{Leading order}
\label{subsec-NLSM-leading}

\begin{figure}
  \centering
   \includegraphics[width=4.3cm]{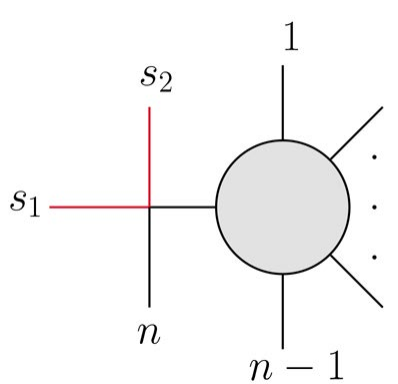}
   ~~~~~~~~~~~~~~~~~~~~
   \includegraphics[width=4.3cm]{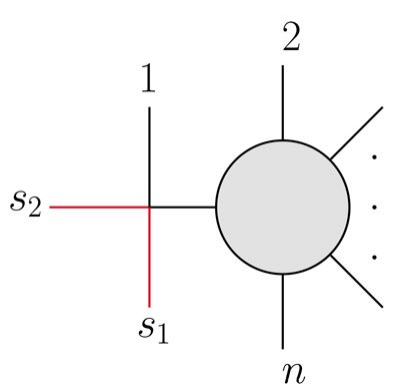} \\
  \caption{Diagrams with divergent propagators in the double-soft limit of NLSM amplitude, where the soft pions represented in red are attached to the quadrivalent vertex in each diagram.}\label{pole-part-NLSM}
\end{figure}

Similar as in the YM case, we expect that two diagrams in Fig.\ref{pole-part-NLSM} contribute to the leading double-soft behavior, since the propagators $1/s_{ns_1s_2}$ and $1/s_{s_1s_21}$, with
\bea
s_{ns_1s_2}&=&2\tau\,k_n\cdot(\hat{k}_{s_1}+\hat{k}_{s_2})+2\tau^2\,\hat{k}_{s_1}\cdot \hat{k}_{s_2}\,,\nn
s_{s_1s_21}&=&2\tau\,k_1\cdot(\hat{k}_{s_1}+\hat{k}_{s_2})+2\tau^2\,\hat{k}_{s_1}\cdot \hat{k}_{s_2}\,,~~\label{s-3pt}
\eea
are divergent in the limit $\tau\to0$. Thus we separate the full amplitude into
\bea
{\cal A}_{\rm NLSM}(s_1,s_2,1,\cdots,n)&=&{{\cal J}_{\rm NLSM}(n,s_1,s_2)\,{\cal J}_{\rm NLSM}(1,\cdots,n-1)\over s_{ns_1s_2}}
+{{\cal J}_{\rm NLSM}(s_1,s_2,1)\,{\cal J}_{\rm NLSM}(2,\cdots,n)\over s_{s_1s_21}}\nn
&&+{\cal R}\,,~~\label{divide-ANLSM}
\eea
where ${\cal J}_{\rm NLSM}$ are scalar currents, and ${\cal R}$ denotes contributions from other diagrams.

At the leading order of $\tau$, the first line on the r.h.s of \eref{divide-ANLSM} behaves as
\bea
&&{{\cal J}^{(0)}_{\rm NLSM}(n,s_1,s_2)\,{\cal J}^{(0)}_{\rm NLSM}(1,\cdots,n-1)\over s_{ns_1s_2}}
+{{\cal J}^{(0)}_{\rm NLSM}(s_1,s_2,1)\,{\cal J}^{(0)}_{\rm NLSM}(2,\cdots,n)\over s_{s_1s_21}}\nn
&=&{{\cal A}_{\rm NLSM}(n,s_1,s_2,I_L)\,{\cal A}_{\rm NLSM}(1,\cdots,n-1,I_R)\over s_{ns_1s_2}}
+{{\cal A}_{\rm NLSM}(s_1,s_2,1,I'_L)\,{\cal A}_{\rm NLSM}(I'_R,2,\cdots,n)\over s_{s_1s_21}}\,.~~\label{NL-leading-polepart}
\eea
Similar as in \eref{YM-leading-middle}, all ${\cal A}_{\rm NLSM}$ in numerators are on-shell amplitudes. Momentum conservation implies
the relation
\bea
&&-k_{I_L}=k_{I_R}=k_n+k_{s_1}+k_{s_2}=k_n+\tau\,(\hat{k}_{s_1}+\hat{k}_{s_2})\,,\nn
&&-k_{I'_L}=k_{I'_R}=k_1+k_{s_1}+k_{s_2}
=k_1+\tau\,(\hat{k}_{s_1}+\hat{k}_{s_2})\,,~~\label{relation-momen-NL}
\eea
which can be reduced to
\bea
-k^{(0)}_{I_L}=k^{(0)}_{I_R}=k_n\,,~~~~~~~~-k^{(0)}_{I'_L}=k^{(0)}_{I'_R}=k_1\,,~~\label{relation-momen-NL-leading}
\eea
thus all external momenta of each ${\cal A}_{\rm NLSM}$ in \eref{NL-leading-polepart} are on-shell. Notice that when considering the leading order in $\tau$, one should in principle drop the ${\cal O}(\tau^2)$ terms in $1/s_{ns_1s_2}$ and $1/s_{s_1s_21}$ in \eref{NL-leading-polepart}. However, performing this procedure would require expanding the factors $1/s_{ns_1s_2}$ and $1/s_{s_1s_21}$ when dealing with higher-order contributions. We find it more convenient to keep these factors as entire elements without expansion. Consequently, we do not discard the ${\cal O}(\tau^2)$ terms in $1/s_{ns_1s_2}$ and $1/s_{s_1s_21}$ even at the leading order. This merely provides an alternative way of organizing soft behaviors at different orders. This is also the manner of organization used in \cite{Cachazo:2015ksa,Du:2015esa}, which provided the standard double-soft theorems for NLSM amplitudes.

The $4$-point on-shell amplitudes in \eref{NL-leading-polepart} can be determined via bootstrap approach. For any $4$-point amplitude ${\cal A}_{\rm NLSM}(a,b,c,d)$, the mass dimension and the cyclical invariance requires
\bea
{\cal A}_{\rm NLSM}(a,b,c,d)=p\,(s_{ab}+s_{bc})=-p\,s_{ac}\,.
\eea
For latter convenience, we choose it as
\bea
{\cal A}_{\rm NLSM}(a,b,c,d)=-k_a\cdot k_c=-k_b\cdot k_d\,,~~~~\label{NLSM-4pt}
\eea
up to an overall constant. Meanwhile, the momenta configuration in \eref{relation-momen-NL-leading} leads to
\bea
{\cal A}_{\rm NLSM}(1,\cdots,n-1,I_R)={\cal A}_{\rm NLSM}(I'_R,2,\cdots,n)={\cal A}_{\rm NLSM}(1,\cdots,n)\,.~~\label{NLSM-npt}
\eea
Plugging \eref{NLSM-4pt} and \eref{NLSM-npt} into \eref{NL-leading-polepart}, we obtain
\bea
&&{{\cal J}^{(0)}_{\rm NLSM}(n,s_1,s_2)\,{\cal J}^{(0)}_{\rm NLSM}(1,\cdots,n-1)\over s_{ns_1s_2}}
+{{\cal J}^{(0)}_{\rm NLSM}(s_1,s_2,1)\,{\cal J}^{(0)}_{\rm NLSM}(2,\cdots,n)\over s_{s_1s_21}}\nn
&=&\Big[{-\hat{k}_{s_2}\cdot k_n\over \hat{s}_{ns_1s_2}}+{-\hat{k}_{s_1}\cdot k_1\over \hat{s}_{s_1s_21}}\Big]\,{\cal A}_{\rm NLSM}(1,\cdots,n)\,,
~~\label{NLSM-leading-part1}
\eea
where
\bea
\hat{s}_{ns_1s_2}={s_{ns_1s_2}\over\tau}\,,~~~~~~~~\hat{s}_{s_1s_21}={s_{s_1s_21}\over\tau}\,.
\eea

Unlike the YM case, whose leading soft behavior is divergent, we see from \eref{NLSM-leading-part1} that the leading contribution from diagrams in Fig.\ref{pole-part-NLSM} is at the $\tau^0$ order. It means other diagrams can also contribute to the leading soft behavior. That is,
\bea
{\cal A}^{(0)}_{\rm NLSM}(s_1,s_2,1,\cdots,n)=\Big[{-\hat{k}_{s_2}\cdot k_n\over \hat{s}_{ns_1s_2}}+{-\hat{k}_{s_1}\cdot k_1\over \hat{s}_{s_1s_21}}\Big]\,{\cal A}_{\rm NLSM}(1,\cdots,n)+{\cal R}^{(0)}\,.~~\label{ansatz1-NLSM-leading}
\eea
Clearly, ${\cal R}^{(0)}$ cannot be fixed via locality and unitarity. However, we can determine it by exploiting the hidden zero with $(i,j)=(1,n)$.
The hidden zero requires that ${\cal R}^{(0)}$ should cancel the remaining part on the r.h.s of \eref{ansatz1-NLSM-leading}. On the other hand, by assumption ${\cal R}$ contains neither $1/s_{ns_1s_2}$ nor $1/s_{s_1s_21}$. Thus we can write our ansatz as
\bea
{\cal A}^{(0)}_{\rm NLSM}(s_1,s_2,1,\cdots,n)&=&\Big[{-\hat{k}_{s_2}\cdot k_n\over \hat{s}_{ns_1s_2}}+{-\hat{k}_{s_1}\cdot k_1\over \hat{s}_{s_1s_21}}\Big]\,{\cal A}_{\rm NLSM}(1,\cdots,n)+2p\,{\cal A}_{\rm NLSM}(1,\cdots,n)+\W{\cal R}^{(0)}\nn
&=&\Big[{-\hat{k}_{s_2}\cdot k_n+p\,\hat{s}_{ns_1s_2}\over \hat{s}_{ns_1s_2}}+{-\hat{k}_{s_1}\cdot k_1+p\,\hat{s}_{s_1s_21}\over \hat{s}_{s_1s_21}}\Big]\,{\cal A}_{\rm NLSM}(1,\cdots,n)+\W{\cal R}^{(0)}\,,~~\label{ansatz2-NLSM-leading}
\eea
where the arrangement in the last step is to keep the manifest symmetry between the two channels. In the above, $\W{\cal R}^{(0)}$ denotes potential terms in ${\cal R}^{(0)}$ which vanish under the constraint of zero kinematics. Logically, we can use the hidden zero to determine $p$, but not $\W{\cal R}^{(0)}$.

For the choice $(i,j)=(1,n)$, the zero kinematics in \eref{zero-NLSM} reads
\bea
k_{s_1}\cdot k_a=0\,,~~k_{s_2}\cdot k_a=0\,,~~~~{\rm for}~\forall\,a\in\{2,\cdots,n-1\}\,.~~\label{kinematic-forNLSM}
\eea
Based on this condition, there are two important observations. The first one is,
\bea
s_{ns_1s_2}+s_{s_1s_21}&=&2\,(k_1+k_n)\cdot(k_{s_1}+k_{s_2})+4\,k_{s_1}\cdot k_{s_2}\nn
&=&-2\,(k_{s_1}+k_{s_2})\cdot(k_{s_1}+k_{s_2})+4\,k_{s_1}\cdot k_{s_2}\nn
&=&0\,,~~\label{observe-1}
\eea
where the second step uses momentum conservation, as well as the zero kinematics in \eref{kinematic-forNLSM}.
This simple relation is the reason why keeping $1/s_{ns_1s_2}$ and $1/s_{s_1s_21}$ as entire elements without expansion is more convenient for our purpose.
The second observation is,
\bea
(k_{s_1}-k_{s_2})\cdot(k_n+k_1)=-(k_{s_1}-k_{s_2})\cdot(k_{s_1}+k_{s_2})=0\,,~~~\label{observe-1.5}
\eea
where momentum conservation and the condition \eref{kinematic-forNLSM} are again used.
Combining \eref{observe-1} and \eref{observe-1.5} together and dropping ${\cal O}(\tau^2)$ terms, we further obtain
\bea
k_{s_1}\cdot(k_1+k_n)=0\,,~~~~k_{s_2}\cdot(k_1+k_n)=0\,,~~\label{observe-2}
\eea

Using the first observation in \eref{observe-1}, we can organize \eref{ansatz2-NLSM-leading} as
\bea
&&{\cal A}^{(0)}_{\rm NLSM}(s_1,s_2,1,\cdots,n)
\,\xrightarrow[]{\eref{kinematic-forNLSM}}\nn
&&{\hat{k}_{s_2}\cdot k_n-p\,\hat{s}_{ns_1s_2}-\hat{k}_{s_1}\cdot k_1+p\,\hat{s}_{s_1s_21}\over \hat{s}_{s_1s_21}}\,{\cal A}_{\rm NLSM}(1,\cdots,n)+\W{\cal R}^{(0)}\nn
&=&{\hat{k}_{s_2}\cdot k_n-2p\,k_n\cdot(\hat{k}_{s_1}+\hat{k}_{s_2})-\hat{k}_{s_1}\cdot k_1+2p\,k_1\cdot(\hat{k}_{s_1}+\hat{k}_{s_2})+{\cal O}(\tau)\over \hat{s}_{s_1s_21}}\,{\cal A}_{\rm NLSM}(1,\cdots,n)\nn
&&+\W{\cal R}^{(0)}\,.
\eea
From the above formula of ${\cal A}^{(0)}_{\rm NLSM}(s_1,s_2,1,\cdots,n)$, we see that the hidden zero at the leading order requires
\bea
\hat{k}_{s_2}\cdot k_n-2p\,k_n\cdot(\hat{k}_{s_1}+\hat{k}_{s_2})-\hat{k}_{s_1}\cdot k_1+2p\,k_1\cdot(\hat{k}_{s_1}+\hat{k}_{s_2})=0\,.~~\label{eq-NL-leading}
\eea
By employing the observation in \eref{observe-2}, the solution of this equation is found to be $p=1/4$.

After substituting $p=1/4$ into \eref{ansatz2-NLSM-leading} and dropping the ${\cal O}(\tau)$ terms, the leading soft behavior then reads
\bea
{\cal A}^{(0)}_{\rm NLSM}(s_1,s_2,1,\cdots,n)
&=&{\cal S}^{(0)}_{\rm NLSM}(n,s_1,s_2,1)\,{\cal A}_{\rm NLSM}(1,\cdots,n)+\W{\cal R}^{(0)}\,,~~~~\label{NLSM-leading-final}
\eea
where
\bea
{\cal S}^{(0)}_{\rm NLSM}(n,s_1,s_2,1)={(\hat{k}_{s_1}-\hat{k}_{s_2})\cdot k_n\over 2\,\hat{s}_{ns_1s_2}}+{(\hat{k}_{s_2}-\hat{k}_{s_1})\cdot k_1\over 2\,\hat{s}_{s_1s_21}}\,.~~\label{soft-factor-NLleading}
\eea
Comparing it with the standard leading double-soft theorem of NLSM amplitudes in \cite{Cachazo:2015ksa,Du:2015esa}, we see that $\W{\cal R}^{(0)}=0$. That is, we have constructed the entire leading soft behavior by solely utilizing locality, unitarity, and hidden zeros.
In subsection \ref{subsec-NLSM-verify}, we will show that this result (with $\W{\cal R}^{(0)}=0$) satisfies hidden zeros with any choice of $(i,j)$.

Before ending this subsection, we would like to give two remarks. Firstly, the dropped ${\cal O}(\tau)$ term in \eref{NLSM-leading-final} is
\bea
{\cal O}(\tau)=\tau\,\Big[{\hat{k}_{s_1}\cdot\hat{k}_{s_2}\over 2\,\hat{s}_{ns_1s_2}}+{\hat{k}_{s_1}\cdot\hat{k}_{s_2}\over 2\,\hat{s}_{s_1s_21}}\Big]\,.~~\label{subleading-kkterm}
\eea
This term will contribute to the sub-leading soft behavior. Secondly, when considering the hidden zero with $(i,j)=(1,n)$ at higher-order of $\tau$,
one need not expand $1/\hat{s}_{ns_1s_2}$ and $1/\hat{s}_{s_1s_21}$ in \eref{NLSM-leading-final}, since they entirely cancel each other as long as the leading soft behavior is organized as in \eref{NLSM-leading-final}.

\subsection{Sub-leading order}
\label{subsec-NLSM-subleading}

Now we move to the sub-leading order. Using the separation in \eref{divide-ANLSM}, we express the sub-leading soft behavior as follows,
\bea
{\cal A}^{(1)}_{\rm NLSM}(s_1,s_2,1,\cdots,n)&=&{{\cal J}^{(0)}_{\rm NLSM}(n,s_1,s_2)\,{\cal J}^{(1)}_{\rm NLSM}(1,\cdots,n-1)\over s_{ns_1s_2}}
+{{\cal J}^{(0)}_{\rm NLSM}(s_1,s_2,1)\,{\cal J}^{(1)}_{\rm NLSM}(2,\cdots,n)\over s_{s_1s_21}}\nn
&&+{{\cal J}^{(1)}_{\rm NLSM}(n,s_1,s_2)\,{\cal J}^{(0)}_{\rm NLSM}(1,\cdots,n-1)\over s_{ns_1s_2}}
+{{\cal J}^{(1)}_{\rm NLSM}(s_1,s_2,1)\,{\cal J}^{(0)}_{\rm NLSM}(2,\cdots,n)\over s_{s_1s_21}}\nn
&&+{\cal R}^{(1)}\,,~~\label{divide-ANLSM-subleading}
\eea
where
\bea
&&{\cal J}^{(0)}_{\rm NLSM}(n,s_1,s_2)={\cal A}_{\rm NLSM}(n,s_1,s_2,I_L)\big|_{\tau=0}=-k_{s_2}\cdot k_n\,,\nn
&&{\cal J}^{(0)}_{\rm NLSM}(s_1,s_2,1)={\cal A}_{\rm NLSM}(s_1,s_2,1,I'_L)\big|_{\tau=0}=-k_{s_1}\cdot k_1\,,\nn
&&{\cal J}^{(0)}_{\rm NLSM}(1,\cdots,n-1)={\cal J}^{(0)}_{\rm NLSM}(2,\cdots,n)={\cal A}_{\rm NLSM}(1,\cdots,n)\,.~~\label{leadingcurrent-NL}
\eea
Using the form of $4$-point amplitudes in \eref{NLSM-4pt}, we find there is no correction to $4$-point on-shell amplitudes in \eref{leadingcurrent-NL} when replacing the momenta configuration \eref{relation-momen-NL-leading} by \eref{relation-momen-NL}.
On the other hand, such replacement yields the following correction to ${\cal A}_{\rm NLSM}(1,\cdots,n-1,I_R)$ and ${\cal A}_{\rm NLSM}(I'_R,2,\cdots,n)$,
\bea
A^{(1)}_{\rm NLSM}(1,\cdots,n-1,I_R)&=&\tau\,(\hat{k}_{s_1}+\hat{k}_{s_2})\cdot\partial_{k_n}\,{\cal A}_{\rm NLSM}(1,\cdots,n)\,,\nn
A^{(1)}_{\rm NLSM}(I'_R,2,\cdots,n)&=&\tau\,(\hat{k}_{s_1}+\hat{k}_{s_2})\cdot\partial_{k_1}\,{\cal A}_{\rm NLSM}(1,\cdots,n)\,,~~\label{sub-correction-NL}
\eea
therefore
\bea
{\cal A}_{\rm NLSM}(n,s_1,s_2,I_L)\,A^{(1)}_{\rm NLSM}(1,\cdots,n-1,I_R)&\in&{\cal J}^{(0)}_{\rm NLSM}(n,s_1,s_2)\,{\cal J}^{(1)}_{\rm NLSM}(1,\cdots,n-1)\,,\nn
{\cal A}_{\rm NLSM}(s_1,s_2,1,I'_L)\,A^{(1)}_{\rm NLSM}(I'_R,2,\cdots,n)&\in&{\cal J}^{(0)}_{\rm NLSM}(s_1,s_2,1)\,{\cal J}^{(1)}_{\rm NLSM}(2,\cdots,n)\,.~~\label{sub-NL-part1}
\eea

At the leading order, we have found that a piece
\bea
2p\,{\cal A}_{\rm NLSM}(1,\cdots,n)~~\label{2p-part}
\eea
belongs to ${\cal R}$ (see in the first line on the r.h.s of \eref{ansatz2-NLSM-leading}), where $p=1/4$. This part also contributes to the sub-leading soft behavior, since ${\cal A}_{\rm NLSM}(1,\cdots,n)$ may receive sub-leading corrections. To capture such sub-leading contributions, we should understand \eref{2p-part} as the leading contribution of
\bea
p\,\Big[{\cal A}_{\rm NLSM}(1,\cdots,n-1,I_R)+{\cal A}_{\rm NLSM}(I'_R,2,\cdots,n)\Big]\,.~~~\label{generalization-2p}
\eea
This is the most natural generalization of \eref{2p-part}, which is consistent with \eref{ansatz2-NLSM-leading}, since ${\cal A}_{\rm NLSM}(1,\cdots,n)$ in
\bea
{-\hat{k}_{s_2}\cdot k_n\over \hat{s}_{ns_1s_2}}\,{\cal A}_{\rm NLSM}(1,\cdots,n)~~~~{\rm and}~~~~
{-\hat{k}_{s_1}\cdot k_1\over \hat{s}_{s_1s_21}}\,{\cal A}_{\rm NLSM}(1,\cdots,n)\,,
\eea
arise from ${\cal A}_{\rm NLSM}(1,\cdots,n-1,I_R)$ and ${\cal A}_{\rm NLSM}(I'_R,2,\cdots,n)$, respectively.
Therefore, by employing \eref{sub-correction-NL}, we find another piece,
\bea
{\cal P}^{(1)}=p\,\tau\,(\hat{k}_{s_1}+\hat{k}_{s_2})\cdot(\partial_{k_1}+\partial_{k_n})\,{\cal A}_{\rm NLSM}(1,\cdots,n)\,.~~\label{sub-NL-part2}
\eea
which also contributes to the sub-leading soft behavior. Notice that \eref{generalization-2p} is not the only admissible extension of \eref{2p-part}, but the difference between the effects of different extensions can be absorbed into ${\cal R}'$ in \eref{ansatz1-NL-subleading}.

Combining \eref{sub-NL-part1}, \eref{sub-NL-part2} and \eref{subleading-kkterm}, we can rewrite \eref{divide-ANLSM-subleading} as,
\bea
&&{\cal A}^{(1)}_{\rm NLSM}(s_1,s_2,1,\cdots,n)\nn
&=&\tau\,\Big[{(\hat{k}_{s_1}-\hat{k}_{s_2})\cdot k_n\over 2\,\hat{s}_{ns_1s_2}}\,(\hat{k}_{s_1}+\hat{k}_{s_2})\cdot\partial_{k_n}+{(\hat{k}_{s_2}-\hat{k}_{s_1})\cdot k_1\over 2\,\hat{s}_{s_1s_21}}\,(\hat{k}_{s_1}+\hat{k}_{s_2})\cdot\partial_{k_1}\Big]\,{\cal A}_{\rm NLSM}(1,\cdots,n)\nn
&&+\tau\,\Big[{\hat{k}_{s_1}\cdot\hat{k}_{s_2}\over 2\,\hat{s}_{ns_1s_2}}+{\hat{k}_{s_1}\cdot\hat{k}_{s_2}\over 2\,\hat{s}_{s_1s_21}}\Big]\,{\cal A}_{\rm NLSM}(1,\cdots,n)\nn
&&+{\cal R}'\,,~~\label{ansatz1-NL-subleading}
\eea
where the second line is the ${\cal O}(\tau)$ term in \eref{subleading-kkterm} which has been dropped when deriving the leading soft theorem, and
\bea
{\cal R}'={\cal M}+{\cal R}^{(1)}-{\cal P}^{(1)}\,,
\eea
where ${\cal M}$ denotes potential missed terms in the first and second lines on the r.h.s of \eref{divide-ANLSM-subleading} which are not detected in \eref{sub-NL-part1}. As in the YM case, an argument analogous to that after \eref{YM-subleading-middle} shows that ${\cal M}$ and ${\cal R}'$ do not contain the pole at $s_{ns_1s_2}=0$ or at $s_{s_1s_21}=0$.

To determine ${\cal R}'$, we again impose the hidden zero with $(i,j)=(1,n)$. Using the observation \eref{observe-1}, we see that terms in the second line of \eref{ansatz1-NL-subleading} cancel each other. Consequently, the first line on the r.h.s of \eref{ansatz1-NL-subleading} should be canceled by ${\cal R}'$. Since ${\cal R}'$ does not contain $1/s_{ns_1s_2}$ or $1/s_{s_1s_21}$, its form is expected to be
\bea
{\cal R}'\,\xrightarrow[]{\eref{kinematic-forNLSM}}\,\tau\,\Big[h\,(\hat{k}_{s_1}\cdot\partial_{k_n}+\hat{k}_{s_2}\cdot\partial_{k_1})
+q\,(\hat{k}_{s_1}\cdot\partial_{k_1}+\hat{k}_{s_2}\cdot\partial_{k_n})\Big]\,{\cal A}_{\rm NLSM}(1,\cdots,n)\,.~~\label{R'-NLSM}
\eea
Substituting this ansatz into \eref{ansatz1-NL-subleading} and using the observation \eref{observe-1}, we get the equation
\bea
&&\Big[(k_1\cdot \hat{k}_{s_2})(\hat{k}_{s_2}\cdot\partial_{k_1})+(k_n\cdot \hat{k}_{s_2})(\hat{k}_{s_2}\cdot\partial_{k_n})-(k_1\cdot\hat{k}_{s_1})
(\hat{k}_{s_1}\cdot\partial_{k_1})-(k_n\cdot\hat{k}_{s_1})(\hat{k}_{s_1}\cdot\partial_{k_n})\nn
&&+4h\,(k_1\cdot \hat{k}_{s_2})(\hat{k}_{s_2}\cdot\partial_{k_1})-4q\,(k_n\cdot \hat{k}_{s_2})(\hat{k}_{s_2}\cdot\partial_{k_n})+4q\,(k_1\cdot\hat{k}_{s_1})
(\hat{k}_{s_1}\cdot\partial_{k_1})-4h\,(k_n\cdot\hat{k}_{s_1})(\hat{k}_{s_1}\cdot\partial_{k_n})\nn
&&+4h\,(k_1\cdot \hat{k}_{s_1})(\hat{k}_{s_2}\cdot\partial_{k_1})-4h\,(k_n\cdot \hat{k}_{s_2})(\hat{k}_{s_1}\cdot\partial_{k_n})
+4q\,(k_1\cdot \hat{k}_{s_2})(\hat{k}_{s_1}\cdot\partial_{k_1})-4q\,(k_n\cdot \hat{k}_{s_1})(\hat{k}_{s_2}\cdot\partial_{k_n})\Big]\nn
&&{\cal A}_{\rm NLSM}(1,\cdots,n)\,\xrightarrow[]{\eref{kinematic-forNLSM}}\,0\,,
\eea
where the higher-order terms are dropped. It is immediate to observe that the second line in this equation cancels the first line, if $h=-1/4$,
$q=1/4$. Suppose this is the correct solution to the equation, the effects of operators in the third line should vanish. To verify this, we separate
${\cal A}_{\rm NLSM}(1,\cdots,n)$ as
\bea
{\cal A}_{\rm NLSM}(1,\cdots,n)=\sum_{m\geq1}\,\lambda_m\,(k_1\cdot k_n)^m+\cdots\,.~~\label{separate-NLSM}
\eea
The zero kinematics in \eref{kinematic-forNLSM} implies
\bea
&&\hat{k}_{s_1}\cdot\partial_{k_1}\,{\cal A}_{\rm NLSM}(1,\cdots,n)=(\hat{k}_{s_1}\cdot k_n)\,\sum_{m\geq1}\,m\lambda_m\,(k_1\cdot k_n)^{m-1}\,,\nn
&&\hat{k}_{s_1}\cdot\partial_{k_n}\,{\cal A}_{\rm NLSM}(1,\cdots,n)=(\hat{k}_{s_1}\cdot k_1)\,\sum_{m\geq1}\,m\lambda_m\,(k_1\cdot k_n)^{m-1}\,,\nn
&&\hat{k}_{s_2}\cdot\partial_{k_1}\,{\cal A}_{\rm NLSM}(1,\cdots,n)=(\hat{k}_{s_2}\cdot k_n)\,\sum_{m\geq1}\,m\lambda_m\,(k_1\cdot k_n)^{m-1}\,,\nn
&&\hat{k}_{s_2}\cdot\partial_{k_n}\,{\cal A}_{\rm NLSM}(1,\cdots,n)=(\hat{k}_{s_2}\cdot k_1)\,\sum_{m\geq1}\,m\lambda_m\,(k_1\cdot k_n)^{m-1}\,.
\eea
Using the above results, together with the observation \eref{observe-2}, we see that operators in the third line cancel each other if $h=-q$.

Substituting \eref{R'-NLSM} into \eref{ansatz1-NL-subleading}, with the solution $h=-1/4$, $q=1/4$, we ultimately get
\bea
{\cal A}^{(1)}_{\rm NLSM}(s_1,s_2,1,\cdots,n)&=&{\cal S}^{(1)}_{\rm NLSM}(n,s_1,s_2,1)\,{\cal A}_{\rm NLSM}(1,\cdots,n)+\W{\cal R}'\,,~~\label{NLSM-subleading-final}
\eea
where the soft factor is given by
\bea
{\cal S}^{(1)}_{\rm NLSM}(n,s_1,s_2,1)=\tau\,\Big[{k_n\cdot \hat{L}_{s_1s_2}\cdot\partial_{k_n}\over\hat{s}_{ns_1s_2}}+
{k_1\cdot \hat{L}_{s_2s_1}\cdot\partial_{k_1}\over\hat{s}_{s_1s_21}}+{\hat{k}_{s_1}\cdot\hat{k}_{s_2}\over 2\,\hat{s}_{ns_1s_2}}+{\hat{k}_{s_1}\cdot\hat{k}_{s_2}\over 2\,\hat{s}_{s_1s_21}}\Big]\,,~~\label{soft-factor-NLsub}
\eea
with $\hat{L}_{s_1s_2}^{\mu\nu}\equiv\hat{k}^\mu_{s_1}\hat{k}^\nu_{s_2}-\hat{k}^\mu_{s_2}\hat{k}^\nu_{s_1}$ and $\hat{L}_{s_2s_1}^{\mu\nu}\equiv\hat{k}^\mu_{s_2}\hat{k}^\nu_{s_1}-\hat{k}^\mu_{s_1}\hat{k}^\nu_{s_2}$. Similar as in \eref{YM-subleading-final}, $\W{\cal R}'$ denotes potential missing terms in ${\cal R}'$ which vanish under the constraint of zero kinematics.

As shown in \cite{Zhou:2023quv,Zhou:2023vzl,Zhou:2024qjh}, the sub-leading soft factor in \eref{soft-factor-NLsub}
is equivalent to the standard one in \cite{Cachazo:2015ksa,Du:2015esa}. In other words, such comparison implies $\W{R}'=0$. Consequently, the sub-leading double-soft theorem can also be completely determined via locality, unitarity and hidden zeros.

\subsection{Verification}
\label{subsec-NLSM-verify}

As in section \ref{subsec-YM-verify}, we can verify the consistency from two aspects.

We again begin by verifying the consistency with other hidden zeros for different $(i,j)$. Given that the leading-order soft theorem in section \ref{subsec-NLSM-leading} is also determined by a combination of locality, unitarity, and the special hidden zero with $(i,j)=(1,n)$, it is necessary to examine both the leading and sub-leading soft theorems. For the following two types of $(i,j)$,
\bea
&&{\rm Case}~1:~i\in\{1,\cdots,n\}\,,~j\in\{1,\cdots,n\}\,,\nn
&&{\rm Case}~2:~i\in\{s_1,s_2\}\,,~j\in\{2,\cdots,n-1\}\,,
\eea
the previous argument in section \ref{subsec-YM-verify} is entirely valid. The only new situations are $(i,j)=(n,s_2)$ and $(i,j)=(s_1,1)$.

Let us focus on $(i,j)=(n,s_2)$. For this choice, the zero kinematics is given by
\bea
k_{s_1}\cdot k_a=0\,,~~~~{\rm for}~\forall\,a\in\{1,\cdots,n-1\}\,.~~\label{kinematic-for-verify}
\eea
From this kinematic condition, a useful observation is,
\bea
0=k_{s_1}\cdot(k_{s_2}+\sum_{i=1}^n\,k_i)=k_{s_1}\cdot(k_{s_2}+k_n)\,,~~\label{observe-forverify}
\eea
where the first step employs the on-shell condition $k_{s_1}^2=0$ and momentum conservation. As a consequence of this observation, we have
\bea
s_{ns_1s_2}=2\,k_n\cdot k_{s_2}+2\,k_{s_1}\cdot(k_{s_2}+k_n)=2\,k_n\cdot k_{s_2}\,.~~\label{s-1}
\eea
Meanwhile, the zero kinematics \eref{kinematic-for-verify} forces
\bea
s_{s_1s_21}=2\,k_{s_2}\cdot k_1+2\,k_{s_1}\cdot k_{s_2}\,.~~\label{s-2}
\eea
Plugging \eref{s-1} and \eref{s-2} into \eref{soft-factor-NLleading}, and using \eref{kinematic-for-verify} and \eref{observe-forverify},
we find
\bea
{\cal S}^{(0)}_{\rm NLSM}(n,s_1,s_2,1)\,\xrightarrow[]{\eref{kinematic-for-verify}}\,\tau\,\Big[{-\hat{k}_{s_1}\cdot \hat{k}_{s_2}\over 2\,\hat{s}_{ns_1s_2}}
+{-\hat{k}_{s_1}\cdot \hat{k}_{s_2}\over 2\,\hat{s}_{s_1s_21}}\Big]\,.~~\label{leading-sub}
\eea
That is, the leading soft factor vanishes at the leading order in $\tau$, and leaves a non-vanishing contribution at the sub-leading order.

Then we move to the sub-leading order. Terms in \eref{leading-sub} arise from the leading soft factor cancel the third and fourth terms in the sub-leading soft factor \eref{soft-factor-NLsub}. The first operator in \eref{soft-factor-NLsub} behaves as
\bea
\tau\,{k_n\cdot \hat{L}_{s_1s_2}\cdot\partial_{k_n}\over\hat{s}_{ns_1s_2}}\,{\cal A}_{\rm NLSM}(1,\cdots,n)
\,&\xrightarrow[]{\eref{kinematic-for-verify}}&\,\tau\,{(k_n\cdot\hat{k}_{s_1})\,\hat{k}_{s_2}\cdot\partial_{k_n}\over\hat{s}_{ns_1s_2}}\,{\cal A}_{\rm NLSM}(1,\cdots,n)\nn
&=&-\tau^2\,{(\hat{k}_{s_1}\cdot\hat{k}_{s_2})\,\hat{k}_{s_2}\cdot\partial_{k_n}\over\hat{s}_{ns_1s_2}}\,{\cal A}_{\rm NLSM}(1,\cdots,n)\,.~~\label{operator1-verify}
\eea
In the above, the first step uses the observation that the zero kinematics in \eref{kinematic-for-verify} forces
\bea
\hat{k}_{s_1}\cdot\partial_{k_n}\,{\cal A}_{\rm NLSM}(1,\cdots,n)=0\,,
\eea
while the second step uses the observation \eref{observe-forverify}. The result in \eref{operator1-verify} shows that the effect of this operator vanishes at the sub-leading order, since the nonzero contribution is at the $\tau^2$ order. The second operator in \eref{soft-factor-NLsub} behaves as
\bea
\tau\,{k_1\cdot \hat{L}_{s_2s_1}\cdot\partial_{k_1}\over\hat{s}_{s_1s_21}}\,{\cal A}_{\rm NLSM}(1,\cdots,n)
\,&\xrightarrow[]{\eref{kinematic-for-verify}}&\,\tau\,{(k_1\cdot\hat{k}_{s_2})\,\hat{k}_{s_1}\cdot\partial_{k_1}\over\hat{s}_{s_1s_21}}\,{\cal A}_{\rm NLSM}(1,\cdots,n)\nn
&=&-\tau^2\,{(k_1\cdot\hat{k}_{s_2})\,(\hat{k}_{s_1}\cdot\hat{k}_{s_2})\over\hat{s}_{s_1s_21}}\,\Big(\sum_{m\geq1}\,m\la_m\,(k_1\cdot k_n)^{m-1}\Big)\,.
\eea
The first step uses the zero kinematics \eref{kinematic-for-verify} which indicates $\hat{k}_{s_1}\cdot k_1=0$. The second step uses the separation
\eref{separate-NLSM}. The final result then follows from the zero kinematics in \eref{kinematic-for-verify} and the observation \eref{observe-forverify}. Consequently, the effect of this operator also vanishes at the sub-leading order, and thus the verification is finished.

Another case $(i,j)=(s_1,1)$ can be treated in the same way. By combining the verifications across different choices of $(i,j)$, we ultimately arrive at the conclusion that the soft factors in \eref{soft-factor-NLleading} and \eref{soft-factor-NLsub} are consistent with all hidden zeros for any chosen pair $(i,j)$.

Then we check the consistency with momentum conservation. Base on the argument similar as that in section \ref{subsec-YM-verify}, we need to verify the condition
\bea
[{\cal S}^{(1)}_{\rm NLSM}(n,s_1,s_2,1)\,,\,P_n]\,{\cal A}_{\rm NLSM}(1,\cdots,n)=\Big((k_{s_1}+k_{s_2})\cdot V\Big)\,{\cal S}^{(0)}_{\rm NLSM}(n,s_1,s_2,1)\,{\cal A}_{\rm NLSM}(1,\cdots,n)\,.~~\label{verify-NL-momen}
\eea
The third and fourth terms in ${\cal S}^{(1)}_{\rm NLSM}(n,s_1,s_2,1)$ in \eref{soft-factor-NLsub} are not operators and thus obviously commute with $P_n$. Therefore, to examine
the equation \eref{verify-NL-momen}, we only need to calculate
\bea
\Big[{k_n\cdot \hat{L}_{s_1s_2}\cdot\partial_{k_n}\over\hat{s}_{ns_1s_2}}+
{k_1\cdot \hat{L}_{s_2s_1}\cdot\partial_{k_1}\over\hat{s}_{s_1s_21}}\Big]\,P_n\,.~~\label{verify-momen-actP}
\eea
To do this, we rewrite these operators as
\bea
{k_n\cdot \hat{L}_{s_1s_2}\cdot\partial_{k_n}\over\hat{s}_{ns_1s_2}}&=&{1\over 2}\,\Big[{\Big(k_n\cdot(\hat{k}_{s_1}-\hat{k}_{s_2})\Big)\,\Big((\hat{k}_{s_1}+\hat{k}_{s_2})\cdot\partial_{k_n}\Big)\over\hat{s}_{ns_1s_2}}
-{\Big(k_n\cdot(\hat{k}_{s_1}+\hat{k}_{s_2})\Big)\,\Big((\hat{k}_{s_1}-\hat{k}_{s_2})\cdot\partial_{k_n}\Big)\over\hat{s}_{ns_1s_2}}\Big]\nn
&=&{1\over 2}\,\Big[{\Big(k_n\cdot(\hat{k}_{s_1}-\hat{k}_{s_2})\Big)\,\Big((\hat{k}_{s_1}+\hat{k}_{s_2})\cdot\partial_{k_n}\Big)\over\hat{s}_{ns_1s_2}}
-{(\hat{k}_{s_1}-\hat{k}_{s_2})\cdot\partial_{k_n}\over2}\Big]\,,\nn
{k_1\cdot \hat{L}_{s_2s_1}\cdot\partial_{k_1}\over\hat{s}_{s_1s_2}1}&=&{1\over 2}\,\Big[{\Big(k_1\cdot(\hat{k}_{s_2}-\hat{k}_{s_1})\Big)\,\Big((\hat{k}_{s_1}+\hat{k}_{s_2})\cdot\partial_{k_1}\Big)\over\hat{s}_{s_1s_21}}
-{\Big(k_1\cdot(\hat{k}_{s_1}+\hat{k}_{s_2})\Big)\,\Big((\hat{k}_{s_2}-\hat{k}_{s_1})\cdot\partial_{k_1}\Big)\over\hat{s}_{s_1s_21}}\Big]\nn
&=&{1\over 2}\,\Big[{\Big(k_1\cdot(\hat{k}_{s_2}-\hat{k}_{s_1})\Big)\,\Big((\hat{k}_{s_1}+\hat{k}_{s_2})\cdot\partial_{k_1}\Big)\over\hat{s}_{s_1s_21}}
-{(\hat{k}_{s_2}-\hat{k}_{s_1})\cdot\partial_{k_1}\over2}\Big]\,,
\eea
where the higher-order terms have been dropped.
Substituting these into \eref{verify-momen-actP}, we immediately get
\bea
&&\Big[{k_n\cdot \hat{L}_{s_1s_2}\cdot\partial_{k_n}\over\hat{s}_{ns_1s_2}}+
{k_1\cdot \hat{L}_{s_2s_1}\cdot\partial_{k_1}\over\hat{s}_{s_1s_21}}\Big]\,P_n\nn
&=&\Big[{(\hat{k}_{s_1}-\hat{k}_{s_2})\cdot k_n\over 2\hat{s}_{ns_1s_2}}+{(\hat{k}_{s_2}-\hat{k}_{s_1})\cdot k_1\over 2\hat{s}_{s_1s_21}}\Big]\,
(\hat{k}_{s_1}+\hat{k}_{s_2})\cdot V\,,
\eea
which completes the verification.

\section{Conclusion and discussion}
\label{sec-conclu}

In this note, we reconstructed the single-soft theorems of tree YM amplitudes and double-soft theorems of tree NLSM amplitudes, at leading and sub-leading orders, from locality, unitarity and hidden zeros. We determined the parts of the soft factors containing specific poles through locality and unitarity, and completed the remaining parts without these poles by means of hidden zeros. Since the full amplitudes can be constructed from these soft theorems, we concluded that locality, unitarity and hidden zeros completely determine YM and NLSM amplitudes at tree level.

A natural future direction is to generalize the method of this note to other amplitudes which exhibit hidden zeros, such as GR amplitudes, namely, to investigate whether locality, unitarity, and hidden zeros completely determine the soft behaviors of these amplitudes. If one attempts to further extend the class of amplitudes determined by $\{$locality, unitarity, hidden zeros$\}$ to a broader scope, it is necessary to simultaneously investigate which other physical models have amplitudes that also exhibit hidden zeros.

As a starting point for investigating whether locality, unitarity, and hidden zeros can completely determine amplitudes, the conclusion of this paper is positive. However, for reasons that will be described below, the method employed in this note has certain shortcomings that need to be addressed. As we have repeatedly pointed out, the approach in this note cannot logically guarantee the completeness of the soft behaviors. Instead, we confirmed the correctness of the obtained soft behaviors by comparing it with known results. This situation is unsatisfactory. When encountering a new physical model that does not have known soft theorem, unlike YM or NLSM, the method presented in this note cannot clarify whether the tree amplitudes of that model are completely determined by locality, unitarity, and hidden zeros.

A possible resolution to the above issue may come from the self-consistency checks performed in sections \ref{subsec-YM-verify} and \ref{subsec-NLSM-verify}. We argued that the soft factors derived through our method are consistent with all hidden zeros with any choice of $(i,j)$. At the same time, these soft behaviors are evidently also consistent with the factorization behavior at physical poles. This set of self-consistencies constitutes a very strong set of constraints. It might be possible to prove that terms which satisfy all the above conditions yet do not appear in the soft factors cannot exist. We will investigate this possibility in future work.

If the above issue can be resolved in future work, then we will possess a general method for constructing soft behaviors from locality, unitarity, and hidden zeros. As noted in section \ref{sec-intro}, a large variety of work in the direction of scattering amplitudes has developed systematic methods for constructing complete amplitudes based on inverting soft theorems \cite{Elvang:2018dco,Nguyen:2009jk,Boucher-Veronneau:2011rwd,Rodina:2018pcb,Ma:2022qja,Luo:2015tat,Zhou:2022orv,
Du:2024dwm,Zhou:2024qjh,Zhou:2024qwm,Wei:2023yfy,Hu:2023lso}. However, these methods require known soft theorems as input. Combining these methods with the approach for constructing soft theorems presented in this note yields a complete on-shell scheme for constructing tree-level amplitudes from locality, unitarity, and hidden zeros.

\section*{Acknowledgments}

This work is supported by NSFC under Grant No. 11805163.

\bibliographystyle{JHEP}

\bibliography{reference}

\end{document}